\documentstyle[12pt,tighten,aps,cite,floats,psfig,prb,prbbib]{revtex}

\newcommand{\psfigure}[1]{\centerline{\psfig{file=#1,width=9cm}}}

\begin{document}
\title{Coarse Graining of Nonbonded Inter-particle Potentials 
Using Automatic Simplex Optimization to Fit Structural Properties}

\author{Hendrik Meyer\thanks{Present address:
     CNRS -- Institut Charles Sadron, 6, rue Boussingault, 67083
     Strasbourg, France},
   Oliver Biermann, Roland Faller, Dirk Reith, and
   Florian M\"uller-Plathe \\
\em Max-Planck-Institut f\"ur Polymerforschung,
        55021 Mainz, Germany}

\date{May 17, 2000, accepted by JCP July 20, 2000}
\maketitle

\begin{abstract}
We implemented a coarse-graining procedure to construct
mesoscopic models of complex molecules.
The final aim is to obtain better results  on properties depending
on slow modes of the molecules.
Therefore the number of particles considered in molecular dynamics
simulations is reduced while conserving as many properties of the
original substance as possible.
We address the problem of finding nonbonded interaction parameters
which reproduce structural properties from experiment
or atomistic simulations.
The approach consists of optimizing automatically nonbonded parameters
using the simplex algorithm to fit structural properties
like the radial distribution function as target functions.
Moreover, any mix of structural and thermodynamic properties can be included
in the target function.
Different spherically symmetric inter-particle
potentials are discussed.
Besides demonstrating the method for Lennard--Jones liquids,
it is applied to several more complex molecular liquids such
as diphenyl carbonate, tetrahydrofurane, and  monomers of poly(isoprene).
\end{abstract}





\section{Introduction}
\begin{figure}
\begin{center}
\psfigure{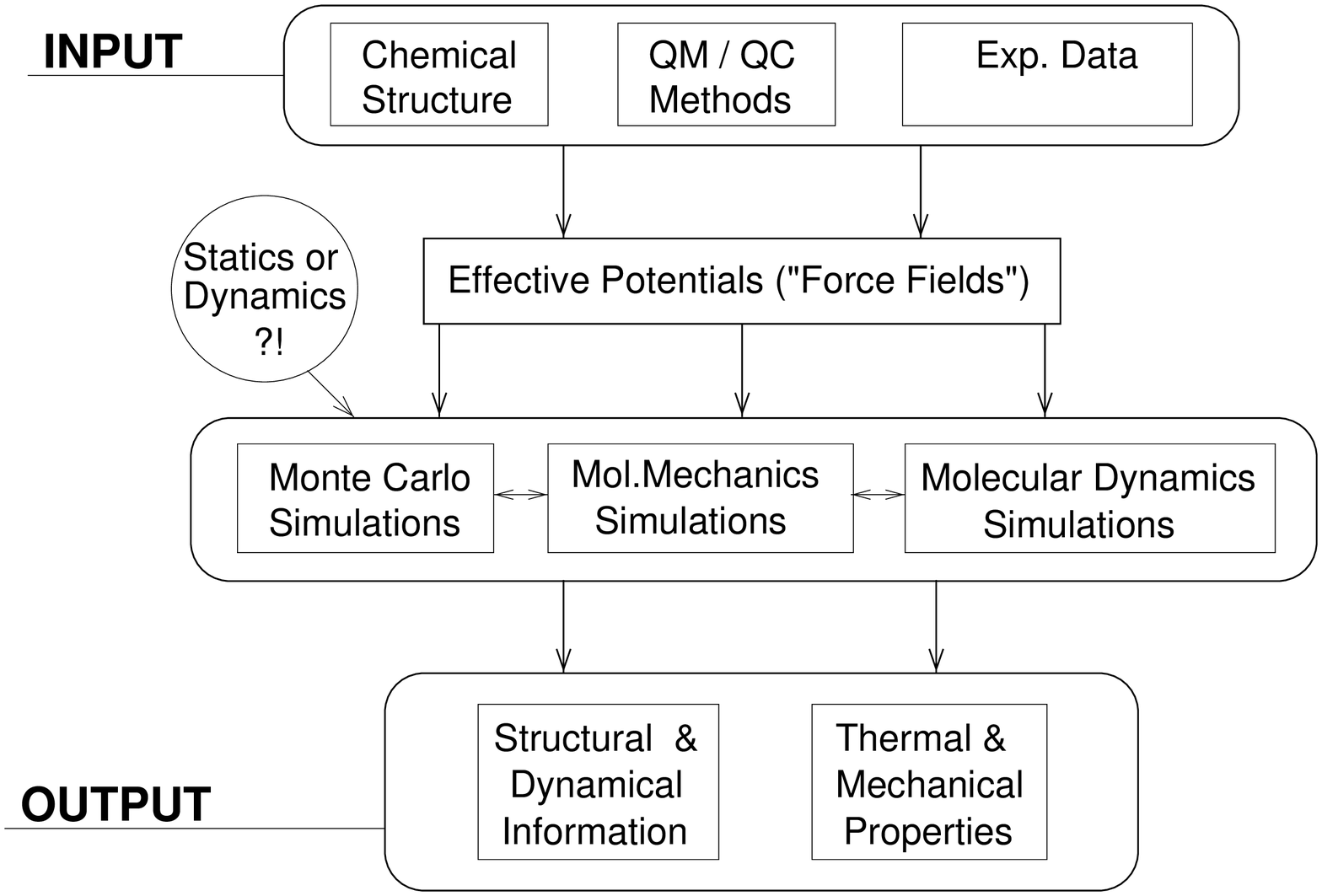}
\end{center}
\caption[Computational path to new information.]{}
\label{comp-path-all}
\end{figure}

Many macroscopic material properties can only be understood by a
thorough investigation of the microscopic details. To obtain
information on a macroscopic level from microscopic input, computer
simulation is, in principle, a proper way. The approach is illustrated
in Fig.~\ref{comp-path-all}.
Taking all available information (e.g.\ quantum chemistry
calculations or experimental data) one builds a force field: a
mathematical function playing the role of an effective potential energy
which is the basis of all molecular simulations. Depending on the specific
question one is interested in (e.g.\ structural or thermodynamic 
properties, statics or dynamics), one chooses the appropriate method 
(e.g.\ Monte Carlo (MC) or molecular dynamics (MD)~\cite{All87}) to get
meaningful results.
This paper discusses nonbonded interaction parameters and how to obtain
them in an efficient way, which represents one important step in
the multiscale modeling of materials. The new idea is to optimize
the simulation parameters with respect to structural distribution
functions.

Our main interest are macromolecular materials where computer
simulations encounter two general problems: on the one hand,
to cover a wide range of length and time scales, and
on the other hand the force field development. 
Firstly, it is almost always necessary to span a wide range of length
and time scales in polymer simulations.  Atomic bonds on the
{\AA}ngstrom-scale are of interest as well as the diffusion of whole chains
of typical extensions of the order of some 100 \AA. Bond vibration times are
of the order of $10^{-13}$ s while the decorrelation time of end-to-end
vectors in entangled polymer melts is of the order of microseconds.  It
is therefore not feasible to perform full-detail atomistic molecular
dynamics simulations to obtain macroscopic bulk properties of
polymers. The detailed treatment of the fast modes would slow down the
run time so strongly that the slow modes can not reach equilibrium.
In addition, the atomistic details sometimes obscure the interesting
properties.
Secondly, it is not trivial to find an appropriate force field for the
investigation of specific properties, no matter if we deal with the
atomistic or mesoscopic level. It was not until recently, that the brute
force ``trial-and-error'' approach was the only choice: repeatedly
guessing a set of parameters, one tries to reproduce some well-known
properties of the system.\cite{mplathe97}
  Modeling the intermolecular interactions by
Lennard-Jones (LJ) potentials, thermodynamic properties like the density
(mostly influenced by the LJ $\sigma$) or the heat of vaporization
(mostly influenced by the LJ $\varepsilon$) are common test observables.
Since it is {\it a priori} unknown how physical properties depend on
some model parameters, much nursing and human intuition is required to
generate a force field by hand.  In this contribution, we try to tackle
both problems.  The development of coarse-grained potentials as well as
the automatic force field refinement procedure proper.

Targeting the first problem, some coarse graining (CG) approaches have
recently been developed to side-step the scaling
problem.\cite{BaEA00aps} They are all similar in spirit: by some
method, a polymer will be mapped onto a less detailed (`coarser')
level. Thus, by integrating out the fast degrees of freedom on the
atomistic scale, the computational task of relaxing the slow degrees
of freedom becomes feasible. Greater time and length scale properties
become accessible and, hence, more information of the
entropy-dominated mesoscale regime can be obtained to derive
macroscopic physical effects.  Technically, the various CG methods are
very different. Some examples are the `time
coarse graining',\cite{forrest95} lattice approaches like the bond
fluctuation model\cite{carmesin88,paul91,baschnagel91} or the high
coordination lattice.\cite{doruker97,doruker99a} On a larger length
scale, dissipative particle dynamics (DPD) and smoothed particle
dynamics (SPS) are frequently used to tackle hydrodynamic
problems.\cite{groot97,espanol97} 
%
For our purposes, an ideal CG method would have the following features:
First, it would start from an atomistic, full-detail polymer model
(typical length scale of bond lengths: a few \AA ngstroms).  Second, a
suitable mapping procedure to a meso-level (typical length scale of one
or a few monomers: some 10 \AA) in continuous space would be needed. It
would take away some atomistic details of the polymer but not its identity.
Therefore, third, it would allow for a unique remapping to the
full-detail level at any stage of the simulation. The whole procedure
would have the advantage that the entropy-driven modes on larger scales
and the energy-driven modes on very local scales are
treated separately, reducing CPU time and loads of disk space
filled with unwanted data on the local structure.  Such a procedure was
successfully implemented for polycarbonate (PC)
melts.\cite{tschoep98a,tschoep98b} It was tested for three different PC
modifications and the resulting microscopic structure compared well to
neutron scattering data.\cite{eilhard99}

Targeting the second problem of force field development, automatic
methods have recently been developed to reduce the human effort
during the optimization procedure. Norrby and Liljefors~\cite{norrby98}
applied numerical Newton-Raphson and simplex algorithms to fit
simultaneously experimental and quantum chemical data for ethane in the
MacroModel software package.  A combination of genetic algorithms and
neural networks was successfully used to construct a force field for
tripod metal templates (`tripodM').\cite{hunger99} The network is
trained on a data base containing experimental structural information.
Another early idea was to use a modified Berendsen weak-coupling method
for automatic parameter adjustment:\cite{muellerplathe95a} By coupling
the time derivative of a force field parameter to the deviation of the
actual value of the target observable, the method could be successfully
established in the case of liquid mercury and SPC water. However, its
applicability is limited to cases in which there is a monotonic
relationship between force field parameter and observable. Della~Valle
et al.~\cite{dellavalle99} use a simplex method to optimize the
intermolecular potential of MD simulations automatically from ab initio
computer simulations of hydrogen fluoride in the liquid phase. The fit
is made against the ab initio potential energy of the dimer.
 We also applied a simplex optimization scheme in MD simulations:
Ref.~\citen{faller99} used thermodynamic properties (density and heat of
vaporization) to derive force fields for atomistic simulations.
In the context of scattering experiments, some ideas similar to those
developed in this work are applied for interpreting experimental data,
e.g. structure functions from neutron scattering, from which one aims to
derive the atomic distributions (so-called reverse MC\cite{rmc} and
empirical potential MC\cite{Sop96cp,Sop97jpcm}).

Historically, force fields have rarely been constructed by automatic
methods so as to reproduce larger-scale structural properties.  Typically,
parameters are chosen by fitting to details such as bond lengths or
angles.  In this work, we select an observable property
which characterises the overall structure and then to automatically
optimize a set of parameters. We concentrate on the
problem of finding an inter-particle potential energy function which
yields a desired radial distribution function (RDF) over the course of
a simulation. The desired RDF is one that is deemed more reliable,
typically from experiment or a higher level calculation such as an
atomistic simulation. Furthermore, we are particularly interested in
exploring the limits of spherical, non-directional interaction
functions since these are fastest to calculate. Fits are performed
using a simplex method, previously used to develop atomistic force
field parameters.\cite{faller99}
%
%
It turns out that much thought has to be
given to the functional form of the inter-particle potential. These
technical details are explained in section II.  As a test case, we
attempt to reproduce the RDF of a Lennard-Jones liquid with various
functional forms of the test potential (section III).
In section IV, we discuss the application of the method to several
real complex liquids which show some variation in chemical structure.
Note that in this paper, we restrict ourselves to the nonbonded part of
force fields and its optimization procedure using low molecular weight
liquids as test systems. Additional aspects which arise due to the
connectivity of large macromolecules will be addressed in a separate
publication.\cite{cgpaper2}

\section{Description of the method}

We are dealing with a so-called inverse problem:
search for an inter-particle potential which reproduces a given
radial distribution function used as the target
function.
To do this automatically, we need three technical ingredients:
(i) specification of the functional form of the
potential and the parameters to be optimized,
(ii) evaluation of the quality of the test potential, 
i.e.\ description of a merit function defining a hyper surface
in the parameter space which has to be minimized, and
(iii) rules to modify the test potential to obtain a better fit
of the target function.
While the latter two points are rather technical and easy to
implement, the specification of the functional form of the 
inter-particle potential turns out to be of utmost importance.

Before going into details, we want to note some practical
constraints: The evaluation of each test potential is quite expensive,
since it requires the simulation of a test system ensuring equilibration
and sufficient statistics for the evaluation of the target quantities.
We have to avoid scanning all parameter space to locate the optimum
potential and instead use an automatic optimization algorithm.  It is
practically impossible to have analytical or reliable
numerical derivatives of our merit function. This excludes many
optimization methods leaving us with the simplex algorithm as one of
the most basic algorithms, which is robust and easy to implement.

\subsection{Choice of the nonbonded potential}

A first brute-force approach
would be to give a general expansion of the nonbonded interaction
potential. The coefficients would then be found by the optimization
algorithm.  However, this straight-forward approach fails in practice,
since there are often too many parameters to optimize. In addition, there
is usually no easy functional dependence between parameters and
target function; the optimization algorithm will often be trapped in local
minima.  This means that one needs physical intuition for
specifying an inter-particle potential in such a way that the parameters
control the potential in a meaningful way.
Besides this physical aspect, there are some more technical points
determining the choice of a potential in computer simulations.  First,
it should be easy and fast to evaluate, second, it
should have a simple analytical form, being continuous and
differentiable. Finally, it is also desirable that it has a physical
interpretation with some theoretical understanding.

Because of technical simplicity, we chose a spherically symmetric potential.
Using non-spherical, directional potentials (like ellipsoids) 
leads to an unwanted increase in computer time and program complexity
without major improvements.
If the molecule under consideration is actually aspherical, it is often
cheaper to model it by two or three spheres than
to introduce one ellipsoid. The experience with the systems reported
in this work shows that spheres can manage most problems.
In addition, a case study for polycarbonates comparing spherical versus
ellipsoidal beads shows that some results are even better reproduced
with spheres.\cite{HaKr99}

Standard spherical potentials used in computer simulations are the 
Lennard-Jones 6--12 potential
\begin{equation}\label{e:potlj}
V_{\rm LJ6-12}(r) = 4\epsilon \left( \left({\sigma\over r}\right)^{12} -
                 \left({\sigma\over r}\right)^{6} \right)  \;,
\end{equation}
or the Buckingham exp--6 potential:
\begin{equation}\label{e:potexp6}
V_{\rm exp-6}(r) = A \, \exp\left(-B{r}\right) -
                 \left({\sigma\over r}\right)^{6}  \;.
\end{equation}
The only term with a physical justification 
is the attractive dispersion term in $1/r^6$ resulting from the
interaction of induced dipoles.
There is also some justification for the exponential repulsion
coming from the limiting behavior of atomic wave functions at large distances.
In general, the repulsive part has no physical justification except that
the potential has to be strongly repulsive at short distances due to the Pauli
exclusion principle.
For modeling larger molecules or monomers of polymer chains
one needs, in most cases, a potential whose repulsive part is softer than
the standard LJ6--12 potential.
For example, one might replace the exponent 12 in the LJ potential by a
smaller integer (still greater than 6) to model a softer core.
However, such a simple two-term potential appears to be not sufficient
if one wants to reproduce more refined structures of RDFs as we found in 
our studies.
We therefore turn to the following general ansatz for the nonbonded potential:
\begin{equation} \label{e:potraw}
V(r) = {a_6 \over r^6} +{a_8 \over r^8} +{a_9 \over r^9} 
 +{a_{10} \over r^{10}} +
  {a_{12} \over r^{12}} + A \, \exp(-Br)  \;.
\end{equation}
This form includes the standard model potentials (\ref{e:potlj})
and (\ref{e:potexp6}), but one has more degrees of freedom allowing to
create softer potentials while keeping the $1/r^6$ term as the
longest-range interaction.
Note that in none of the problems studied here, we use all of Eq.\
(\ref{e:potraw}) simultaneously.

\begin{figure}
\psfigure{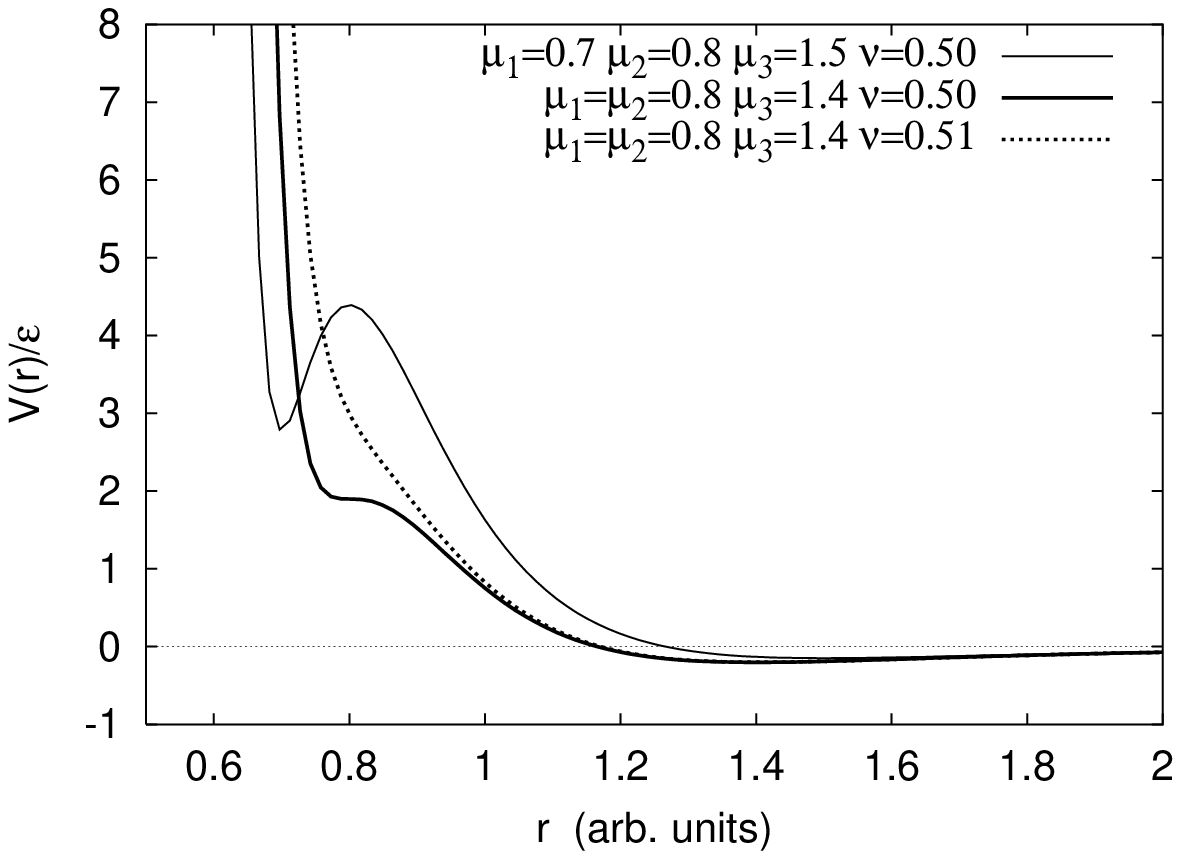}
\caption[Examples of the 6-8-10-12-potential:
Thin continuous line: $\mu_{1}< \mu_{2}< \mu_{3}$ and $\nu=0.5$ yield two
minima;
thick continuous line: $\mu_{1}= \mu_{2}< \mu_{3}$ and $\nu=0.5$ 
yields a saddle point;
thick broken line: $\nu>0.5$ smears out the first minimum or saddle
point.]{}
\label{f:potmm}
\end{figure}

For our kind of problem, it is difficult to adjust more than 4--5 free
parameters at once in reasonable computer time because each set of
parameters needs a full MD simulation run.\cite{faller99} Hence, a
choice of some parameters is necessary.  It turns out that the form of
Eq.\ (\ref{e:potraw}) is not suitable for optimization by varying
directly the parameters $a_n$.  It is advantageous to rewrite the
potential $V(r)$ with parameters varying it in a defined way, with
meaning attributed to at least some of them.  The LJ6-12 potential, for
example, is usually written in the form (\ref{e:potlj}) with $\sigma$
being the approximate diameter of the excluded volume and $\epsilon$ the
depth of the potential minimum.  A similar interpretation can be found
with other two-term potentials.  It is not so straight-forward to
rewrite expansions with more terms into a useful form.  We propose here
a four term potential 6-8-10-12 which was quite successful for the
description of very broad RDFs:
\begin{equation}\label{e:potmm}
V(r) = - {\epsilon \over r^{12}} \left( r^{6}
         - 0.75 \, (\mu_{1}^{2}+\mu_{2}^{2}+\mu_{3}^{2}) \, r^{4}
         + 0.6  \, (\mu_{1}^{2}\mu_{2}^{2}+\mu_{2}^{2}\mu_{3}^{2}
                 +\mu_{1}^{2}\mu_{3}^{2}) \, r^{2} 
         - \nu \, \mu_{1}^{2}\mu_{2}^{2}\mu_{3}^{2}  \right)   \;,
\end{equation}
where the coefficients $a_6, a_8, a_{10},$ and $a_{12}$ of
Eq.~(\ref{e:potraw}) have been rewritten in terms of
new variables $\epsilon, \mu_1, \mu_2, \mu_3$, and $\nu$. 
This parameterization was constructed such that
for $\nu=0.5$ the three parameters $\mu_i$
give the location of the maxima and minima of the potential
(see the appendix for the details of the derivation).
For $\epsilon>0$ and $0< \mu_{1}< \mu_{2}< \mu_{3}$,
 $\mu_{1}$ and  $\mu_{3}$ are the locations of minima and
$\mu_{2}$ is the location of the maximum in-between.
For $\mu_{1}= \mu_{2}$ or $\mu_{2}= \mu_{3}$, there is a saddle point
at $\mu_2$. Some examples are sketched in Fig.~\ref{f:potmm}.
In most cases, one does not want to have two minima in the potential
but only one minimum which is broader than in the LJ6-12 case.
This can be achieved by increasing the coefficient $\nu$
of the repulsive $1/r^{12}$ term. A value slightly above 0.5
is usually sufficient as one can see in the discussion of the results.
With this approach, one can obtain two different repulsive regimes:
a strongly repulsive part below $\approx\mu_1$ which marks the
hard core and a weakly repulsive regime representing a softer shell
(see for example Fig.~\ref{f:rdfiso} in Sec. IV A).

\begin{figure}
\psfigure{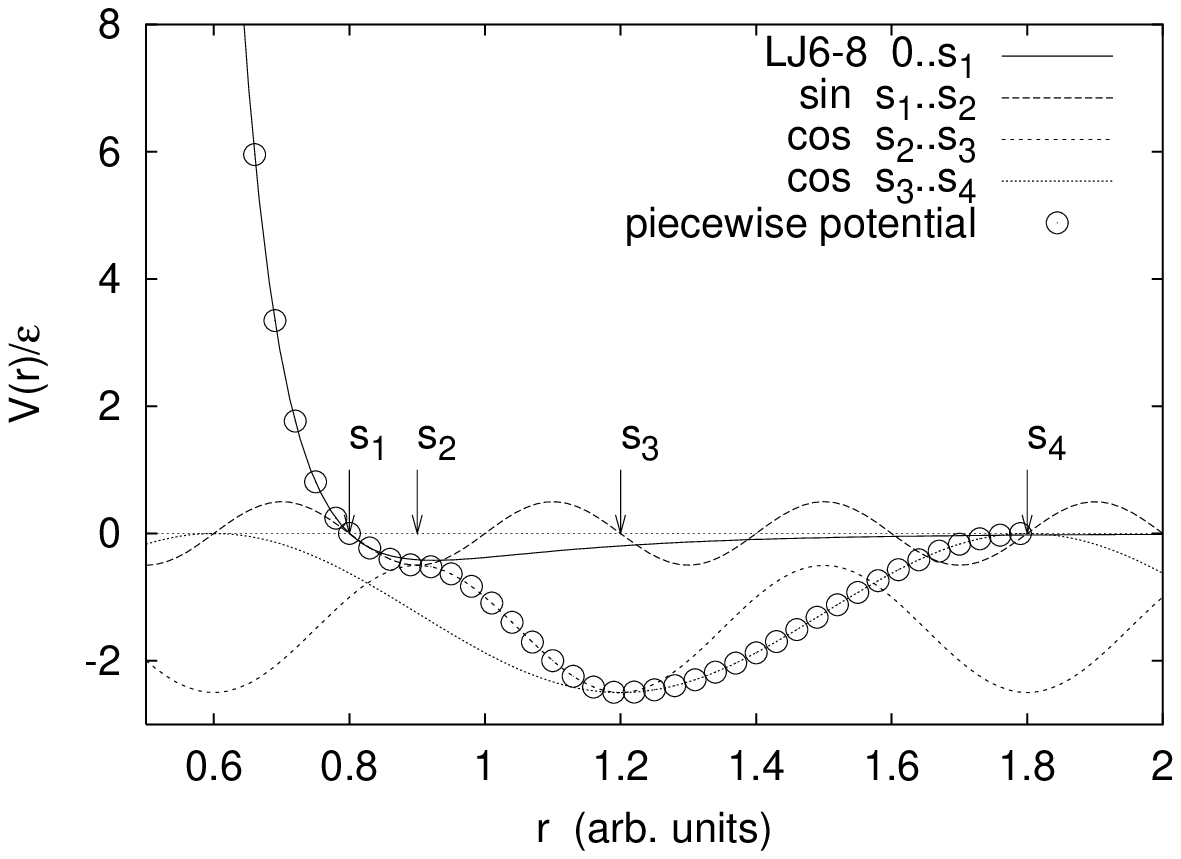}
\caption[Illustration of the piecewise defined potential Eq.\ 
  (\protect\ref{e:potpiece}): a repulsive core is modeled by a
 LJ6-8 potential, the minimum consists of several cosine
 functions.]{}
\label{f:potpiece}
\end{figure}

At first sight, the analytical form (\ref{e:potraw}) seems to fulfill
all needs.  Unfortunately, this is not always the case as we will show during
the application to real molecules.  For systems where the first shell
peak of the RDF has some complicated structure, a potential constructed
of piecewise different functions is more useful
(see Fig.~\ref{f:potpiece}):
\begin{equation}\label{e:potpiece}
V(r) = \left\{\begin{array}{ll}
          \epsilon_1\,\left( \left({\sigma_1 \over r}\right)^8 
                             - \left({\sigma_1 \over r}\right)^6 \right) 
            &  r < \sigma_1 \\
          \epsilon_2\,\left(\sin{(\sigma_1-r) \pi \over (\sigma_2-\sigma_1)2 }
                       \right)
            &  \sigma_1 \le r < \sigma_2 \\
          \epsilon_3\,\left(\cos{(r-\sigma_2) \pi \over \sigma_3-\sigma_2 }
                      - 1 \right) - \epsilon_2
            &  \sigma_2 \le r < \sigma_3 \\
          \epsilon_4\,\left(-\cos{(r-\sigma_3) \pi \over \sigma_4-\sigma_3 }
                      - 1 \right) - \epsilon_2
            &  \sigma_3 \le r < \sigma_4\equiv r_{\rm cut} \\
       \end{array} \right.
\end{equation}
The first part is a repulsive LJ6--8 potential up to $\sigma_1$ to model
the core of excluded volume of the particle. The second and the third
part are again repulsive,  with force minima at $\sigma_2$ and
$\sigma_3$.
This yields an RDF with a first peak or a shoulder at $\sigma_2$ and
a second one above $\sigma_3$.
One might omit the second part by setting $\epsilon_2=0$ and
$\sigma_1=\sigma_2$.
 Whether there appear two peaks
or one peak and a shoulder depends on $\epsilon_3$ and
the size of the interval $[\sigma_2\dots\sigma_3]$.
The fourth term is optional and provides an attractive part.
This influences usually the width of the first-shell peak
of the RDF.
The whole potential was shifted such that $V(r_{\rm cut})=0$.
The cosine potentials have the nice property that their derivative
(the force) vanishes at $r=\sigma_2$ and $r=r_{\rm cut}$.
This ensures that the forces are continuous at the cutoff distance.
However, the force is discontinuous at $\sigma_1$.

\subsection{The merit function}

For the automatic optimization, a prescription is needed to get a number
measuring the quality of the radial distribution function (RDF) of a
test potential.  We use the squared difference between the
simulated RDF $g_s(r)$ and the target RDF $g(r)$ integrated over an
appropriately chosen interval $[r_{\rm min}\dots r_{\rm max}]$,
optionally weighted with some weighting function $w(r)$
\begin{equation} \label{e:fi}
f = \int_{r_{\rm min}}^{r_{\rm max}} w(r) 
            \left(g(r) - g_s(r)\right)^2  dr \;.
\end{equation}
This merit function $f$ is zero for perfect agreement and the more
positive the worse the agreement.  Eq.~(\ref{e:fi}) defines the
hypersurface in the parameter space which has to be minimized. (In
optimization theory one often speaks of a potential surface -- we avoid
this term since the name `potential' is used for the input parameter.)
The weighting function $w(r)$ in Eq.~(\ref{e:fi}) was chosen to be a
decaying exponential $w(r)=\exp(-r/\sigma)$ to penalize stronger
deviations at small separation ($\sigma$ being the typical unit length
of the system). However, after the experience of this work we expect
that this weighting is not so important since contributions in the
excluded volume region are naturally large if the first shell peak is at
the wrong position.  Note also, that this merit function cannot give an
absolute quality value since there is no unique way of normalization.
It depends always on the given problem what threshold value separates
``good'' from ``bad'' quality.

One can include other quantities as the deviation from
a target pressure or other thermodynamic quantities as
additional terms in (\ref{e:fi}).
In principle, the RDF contains all information about the system.
However, there are many reasons like finite size, truncation
of potentials, or simply convergence behavior
which make it useful to enforce thermodynamic properties
independently.

\subsection{The simplex algorithm}

\begin{figure}
\psfigure{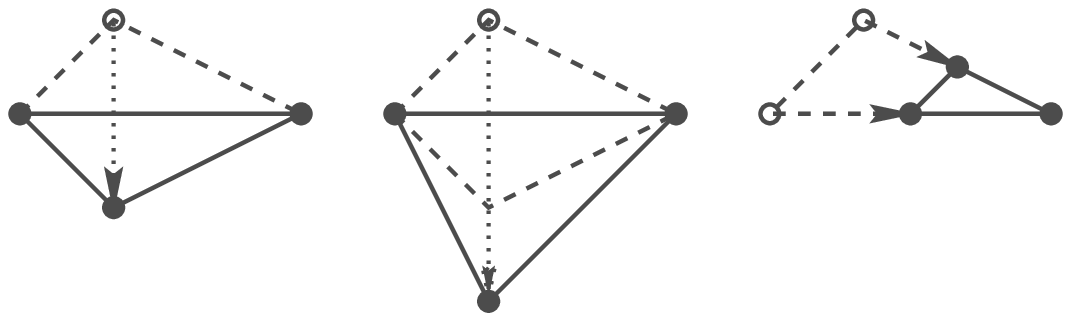}
\caption[Two-dimensional illustration of the three standard simplex moves
reflection, expansion and contraction (see text).]{}
\label{f:simplexmoves}
\end{figure}

The simplex algorithm is a multi-dimensional optimization
procedure.\cite{NumRec} It allows for maximum generality in the merit
function as it does not use derivatives.  In our case this is useful,
since the actual function is not even known.
As there exist many descriptions of this algorithm, we present only a
brief review. A simplex is a set of $d+1$ points in a $d$-dimensional
parameter space. Any subset with $d$ of these points must be linearly
independent.
One has to set up a starting simplex which should cover approximately
the region where the optimum is expected.
After evaluation of these initial points, the actual simplex
algorithm starts 
(see Fig.~\ref{f:simplexmoves} for a two-dimensional illustration).
All but the worst point (largest merit function value $f$) define a
hyper-plane through which the worst point is reflected.  Depending on
the evaluation of this new point a further expansion into the promising
direction, nothing or a (one-dimensional) contraction to the ``safe''
hyper-plane is performed. If none of these leads to an improvement, a
$d$-dimensional contraction is performed by moving all but the best
point towards the best.  This procedure is iterated until the merit
function of the best point falls below a given threshold value. The
iteration might also be stopped if the simplex becomes too small or
after a user defined maximum of iteration steps is exceeded.  In these
two cases the optimization has failed.  One might restart it with
different starting points.

One function evaluation step consists of an entire MD simulation with
the potential defined by the simplex point in parameter space. This is
quite expensive since one needs some time to reach an equilibrium
configuration with respect to the given potential and a certain time of
simulation in equilibrium to have a sufficiently large statistical basis
for the evaluation of the target quantities.
The examples presented in this paper were run on DEC-Alpha
workstations where one simplex step with 400 CG beads
typically needs one hour of CPU time.

Let us finally mention a technical difficulty appearing in the first
equilibration stage. One has to give a start configuration of particle
positions. For this, we chose the final configuration of a run whose
potential parameters are closest to the new ones.  In most cases, this
reduces the simulation time needed for equilibration. However, if the
repulsive part of the new potential is stiffer than for the potential
with which the start configuration was produced, there can be some
particles which are so close to each other that they are deeply in the
excluded volume region with respect to the new potential. This can lead
to numerical problems which manifest themselves usually by divergence of
the MD integrator.  Therefore we introduced short pre-equilibration runs
with a much shorter time step and an upper bound for the inter-particle
forces to circumvent program failures due to floating point errors.

\section{A test case: Reproducing the structure of a Lennard--Jones liquid}

As a first test case, we performed simulations of a standard LJ6-12
potential with $\epsilon=\sigma=1$, i.e.\ $V(r)$ of Eq.\ 
(\ref{e:potraw}) with $a_6=-4.0$ and $a_{12}=4.0$, all other parameters
being zero. A cutoff distance of $2.52 \sigma$ was imposed for the potential
interaction.  The system contained 400 particles at constant density $\rho=0.85
\sigma^{-3}$ in a cubic box with periodic boundary conditions. The
temperature $T=1$ $\epsilon/k_B$ was maintained by a Langevin-thermostat
with friction constant $\Gamma=0.5$.  From this simulation we get an RDF
which is used as target function in the following.  The simulations
for the optimization process are run under the same conditions except,
of course, that the potential is modified.

\subsection{Buckingham potential}

The purpose of this section is to find parameters of the Buckingham
exp-6 potential which reproduce best the RDF of a LJ6-12 potential with
$\epsilon=\sigma=1$.  The Buckingham exp-6 potential (\ref{e:potexp6})
is obtained from the general form Eq.~(\ref{e:potraw}) by setting
$a_8=a_9=a_{10}=a_{12}=0$.  Using the remaining three parameters
$A,B,a_6$ as simplex parameters may lead to problems because the
potential is extremely sensitive to these parameters. Small changes yield
a potential which is completely unphysical or give problems with
equilibrating the simulation. (Note that the exp-6 potential has an
unphysical asymptotic behavior for $r\rightarrow 0$. During simulation,
this regime is usually never reached since it is separated by an energy
barrier high of many $kT$.  During optimization, however, a set of parameters
might be chosen for which the repulsive barrier is not large enough.)
One can rewrite the exp-6 potential in the form
\begin{equation} \label{e:exppara}
V(r) = \epsilon \left(b_1 \exp(-b_2[r/\sigma]) + [r/\sigma]^{-6} \right)
\end{equation}
to have global scaling parameters $\epsilon$ and $\sigma$
for the energy and the distance respectively. Of course, this
yields a redundant fourth parameter, and in the following, we always set
$\sigma=0.4$.

\begin{figure}
    \psfigure{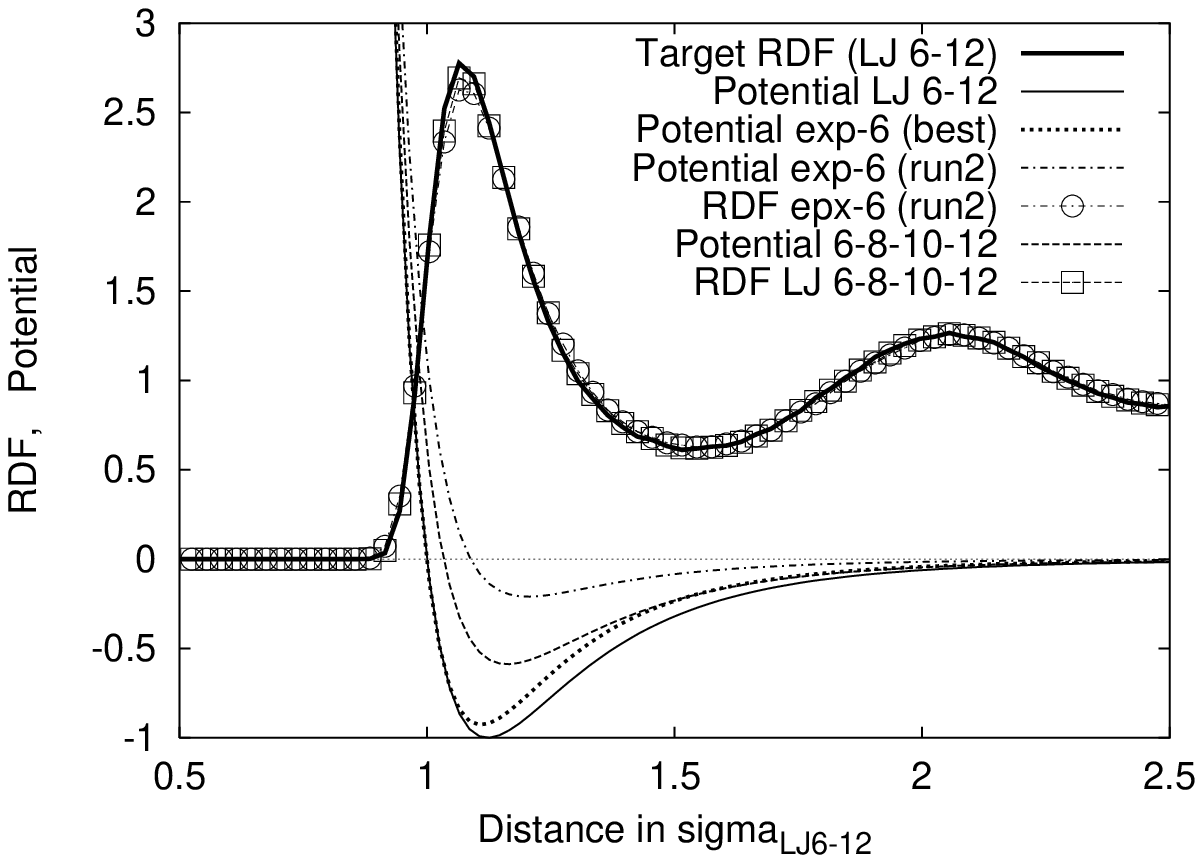}
    \caption[LJ 6-12 liquid approached by 6-8-10-12 potential
      and Buckingham exp-6. Thick continuous line: target RDF of LJ6-12
      liquid with $\epsilon=\sigma=1$.  The best RDF with the Buckingham
      exp-6 potential (dotted line) cannot be distinguished and is not shown.
      Circles represent the RDF of the dot-dashed exp-6 potential named
      `run2'.       Also plotted is an RDF
      (squares) with a 6-8-10-12 potential which is quite similar
      despite a quite different potential.]{}
    \label{f:rdflj}
\end{figure}

A start simplex is constructed by least-squares fitting $V(r)$ to the
LJ6-12 potential in different regions. This yields already points with a
very good agreement of the RDFs, see first block in Table \ref{t:exp6}.
Note that this way of obtaining start parameters is only feasible if one
already has a functional form of the potential and one only searches
for a different parameterization.
%
The integral in the merit function $f$ was evaluated on 70 equidistant bins up
to $r_{\rm max}=2.1$, we chose as a convergence threshold 0.001.
There are some jumps during the first
10 optimization steps, afterwards the algorithm converges very quickly
to merit function values between 0.0015 and 0.0025 and very
small variation in the parameters optimized by the simplex algorithm.
In fact, the statistical precision of our target RDF as well as of
the simulated optimization steps was not sufficient for so small
differences. The best value appeared already after 28 steps (out of
100).
The corresponding RDF cannot be distinguished from the target RDF by
visual inspection.  Interestingly, the potential coincides only in the
repulsive part, being slightly different in the attractive region (see
Fig.~\ref{f:rdflj}).  This reflects the well-known fact, that for dense
liquids, the repulsive part of the potential is the most important for
determining the structure.\cite{weeks71} The pressure is positive and
about 30\% higher than in the LJ6-12 reference system.

To test the capabilities of the simplex algorithm with less perfect
starting points, we chose starting values which fitted the
Buckingham-potential to a LJ6-12 potential with $\epsilon=1/4$ 
(see block (B) in Table \ref{t:exp6}).
After about 20 optimization steps, the optimized parameters showed very
little variation and the optimization was stopped after 45 steps. The
best values according to the merit function are also given in
Table \ref{t:exp6}.
In this case, the simplex algorithm is trapped in a local minimum which
is not as good as before. Nevertheless, if one keeps in mind that the
starting values were completely off, the agreement of the RDFs is
already quite good (see circles in Fig.~\ref{f:rdflj}).
In contrast, the corresponding potential is completely different
with a less pronounced minimum. This means, that the RDF is more 
a result of the constant volume constraint than of the specific
form of the potential.
This would probably be different if one performed simulations in the
NpT ensemble. In this case one should include
the density as a target value in the merit function similarly to
Ref.~\citen{faller99}. Here, the NVT ensemble was chosen to impose
the correct density so that the optimization has only to deal with
the structure.

\subsection{LJ6-9 potential}

\begin{figure}
  \psfigure{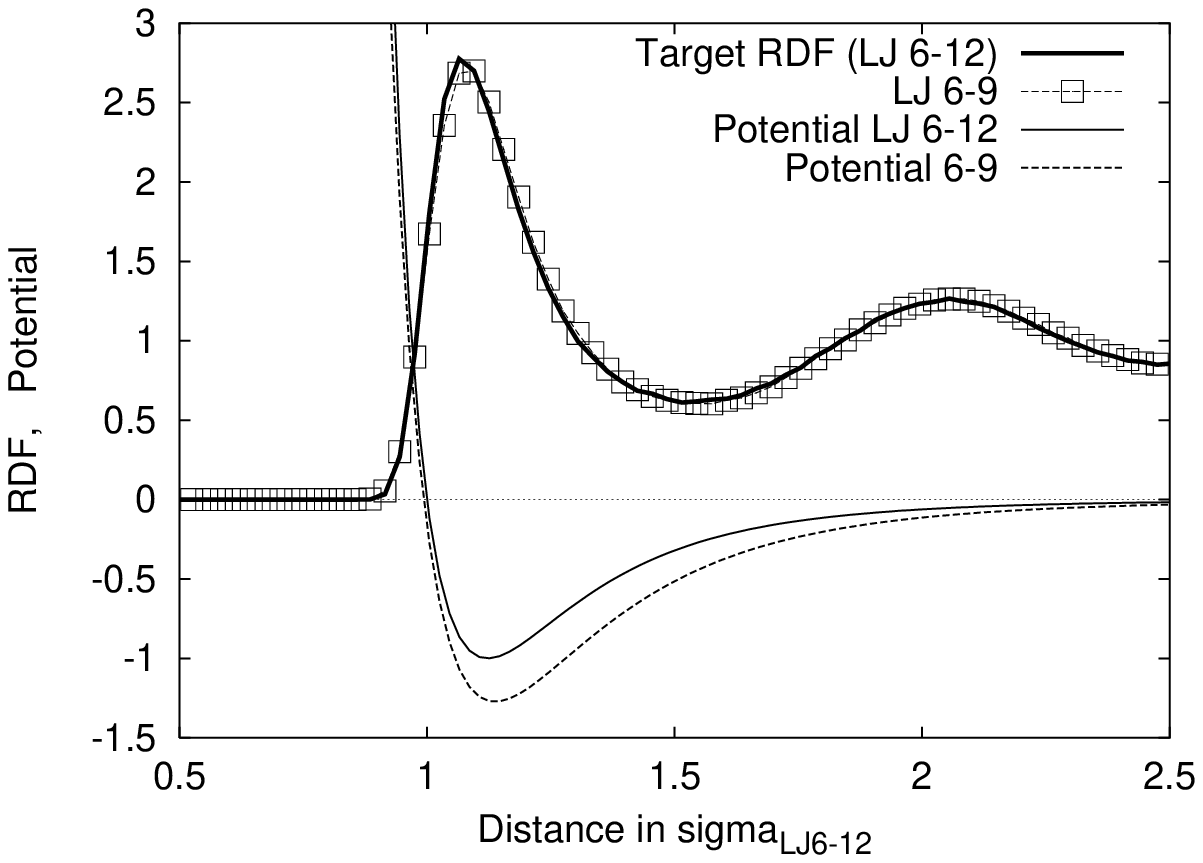}
    \caption[LJ 6-12 liquid modeled by 6-9 potential.
      Thick continuous line: target RDF of LJ6-12 liquid with
      $\epsilon=\sigma=1$;  broken line with symbols: RDF of 6-9
      potential.  Corresponding potentials are drawn with the same
      linestyle as the RDFs they belong to.]{}
 \label{f:lj69}
\end{figure}

\begin{figure}
  \psfigure{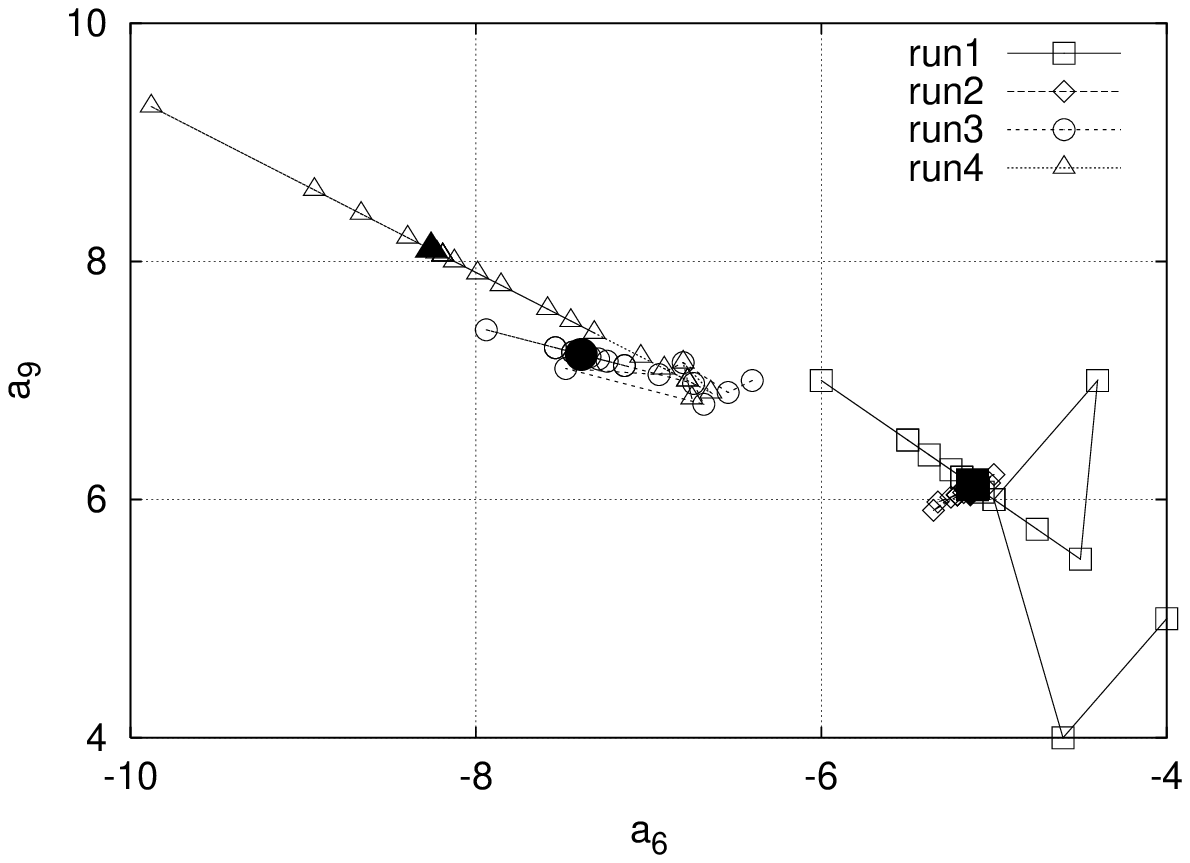}
  \caption[Optimizing an LJ6-9 potential to fit the RDF of a LJ6-12
    liquid: Parameter space ($a_6,a_9$)-plane sampled by the simplex
    algorithm with 4 independent start value sets. The filled symbols
    mark the best point of each run; the numerical values of these
points are given in Table \protect\ref{t:lj69}.]{}
  \label{f:path69}
\end{figure}

Next we try to find out how good one can approach the RDF of the LJ6-12
potential with $\epsilon=\sigma=1$ if we set the repulsive exponent
to 9 (see Fig.~\ref{f:lj69}).
Table \ref{t:lj69} shows the parameters and the merit function value
of a rather bad start guess. The quality does not significantly change
after 20 steps. After 3 restarts an acceptable parameter set
is found in a quite different region of parameter space. Note that
the pressure for the system with this potential is much lower than
in the LJ6-12 reference system.
%
The curves marked 6-9 in Fig.\ \ref{f:lj69}
correspond to it.
Fig.~\ref{f:path69} shows the points in the parameter space which
were visited during the simplex optimization process during the 
4 runs.
(run1) has the largest jumps whereas the following runs show less
variation.  The figure allows us another interesting observation: The
simplex often ends up on a line.  This can happen easily for
2-dimensional optimization runs, when no new point can be found which is
better than the best two points.  It means that the best two points
are on the bottom of a valley and the worst point is reflected from one
side to the other with respect to this valley.  This effect will
increase when the differences in the merit function are more due to
statistical variations than to systematic dependence on the test
potential.

\subsection{LJ6-8-10-12 potential}

We now propose a potential with more free parameters in which the
original potential is contained to see if
the automatic optimization finds back the correct form.
We use the four coefficients $a_{n}$ ($n=6,8,10,12$) as parameter
of the simplex optimization, starting with an arbitrary guess
of a very bad simplex:

After about 30 steps, the simplex is trapped in a local minimum.  The
potential parameters as well as the quality of the RDF do not vary
substantially.  In Fig.~\ref{f:rdflj}, the RDF and the potential \#99
are plotted.  The pressure of this system is about twice as high than in
the LJ6-12 reference system.  The LJ6-8-10-12 potential has a much
shallower minimum than the original LJ6-12 potential. This means that
the RDF must be due in part to the constant-volume condition, just like
in the case of the Buckingham potential.
When one restarts the optimization with starting values
closer to the LJ6-12 values, the success is much better, which is not
so surprising (see block (B) in Table \ref{t:lj681012}). 
In this case, the simplex algorithm did not
start because points 1 and 3 were already below the
threshold of 0.01; the corresponding RDF is indistinguishable
from the target RDF.

\subsection{First summary}

In this section, we reported the first experiences of optimizing an
inter-particle potential to fit structural distribution functions. These
test runs show that our implementation of the automatic simplex
optimization works well and that the optimization with respect to
distribution functions is possible.
Nevertheless, some pitfalls already appeared: The simplex algorithm works
best when the optimum value is inside the starting simplex. In our
implementation, the algorithm is able to find an optimum outside.
However, if the simplex becomes too large, it is not very efficient and
it becomes likely to fall into a local minimum which is not the best
one. One has to be careful with 2-dimensional simplices.

On the physical side, we observe that one can get the RDF
approximately right with completely different potentials. To some
extent, this is due to the fact that most of the RDF is determined by
the repulsive part of the potential. This effect is enforced by the
constant-volume constraint.

\section{Coarse graining of real molecular liquids}

\begin{figure}
\centerline{\psfig{file=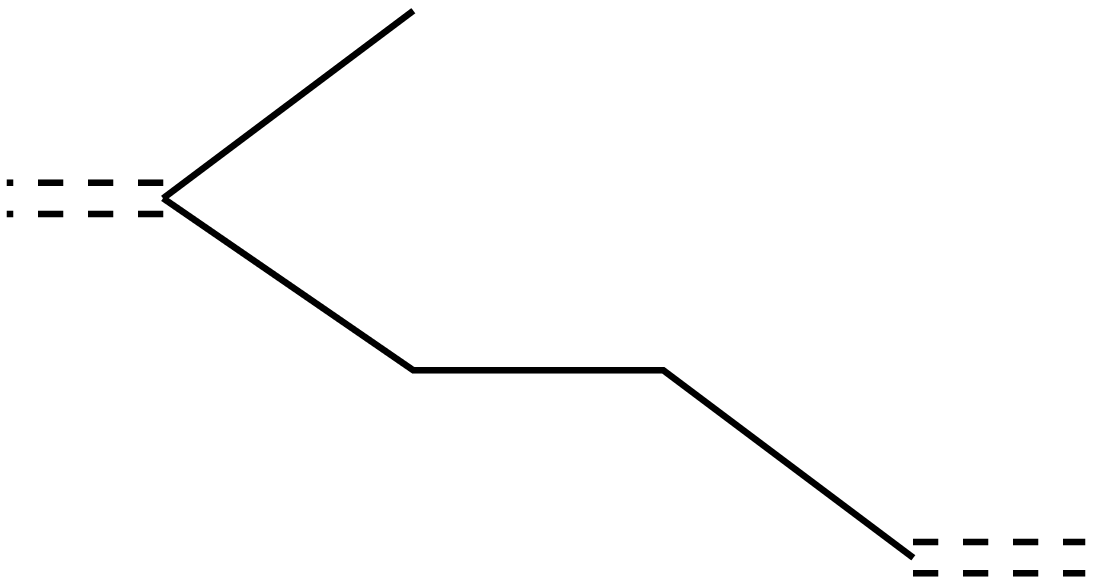,height=2.3cm}\qquad
 \psfig{file=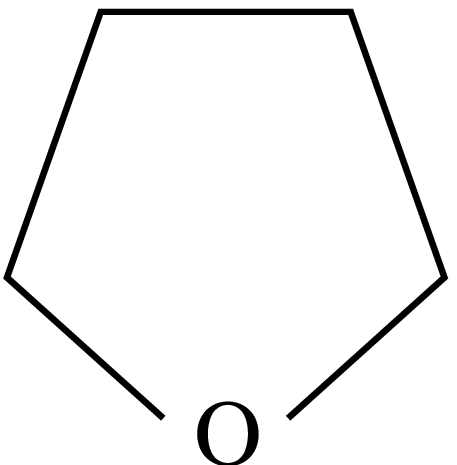,height=2.2cm}\qquad
 \psfig{file=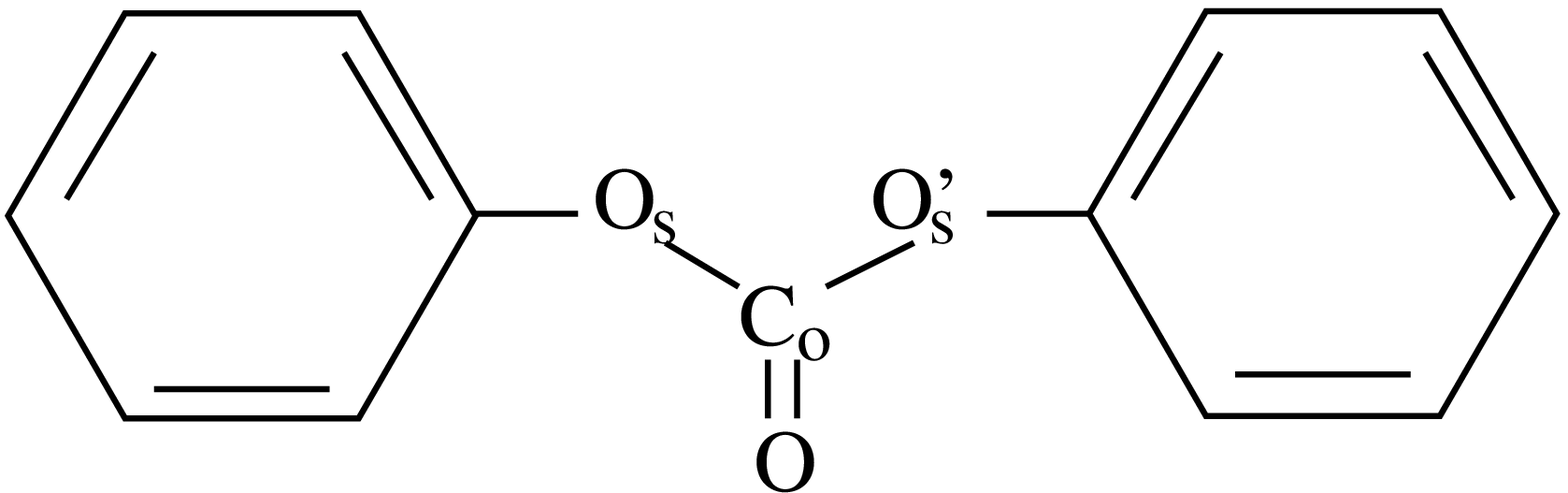,height=2.2cm}}
\caption[Chemical structure of the three molecules used in this section:
 pseudo monomer of poly(isoprene), tetrahydrofurane (THF), and diphenyl
 carbonate (DPC).]{}
\label{f:mols}
\end{figure}

The success of the preceding section is  not too  surprising
since the original system, the Lennard-Jones 6-12 liquid, is more or
less contained in the fitting potential.
We discuss now three real molecular liquids with increasing
complexity; their chemical structures are shown in Fig.~\ref{f:mols}.
 The target function is always the center-of-mass RDF (CM-RDF)
obtained from atomistic simulation data.
The first approach is always to model the whole molecules by one bead
with a spherically symmetric potential.
In the third case of DPC which is very aspherical, we report results
with alternative models comprising two or three beads per molecule.
The systems were chosen because they are important in ongoing research
projects in our laboratory. They also represent some variation in
chemical structure.

\subsection{Isoprene Pseudo Monomers}

\begin{figure}
    \psfigure{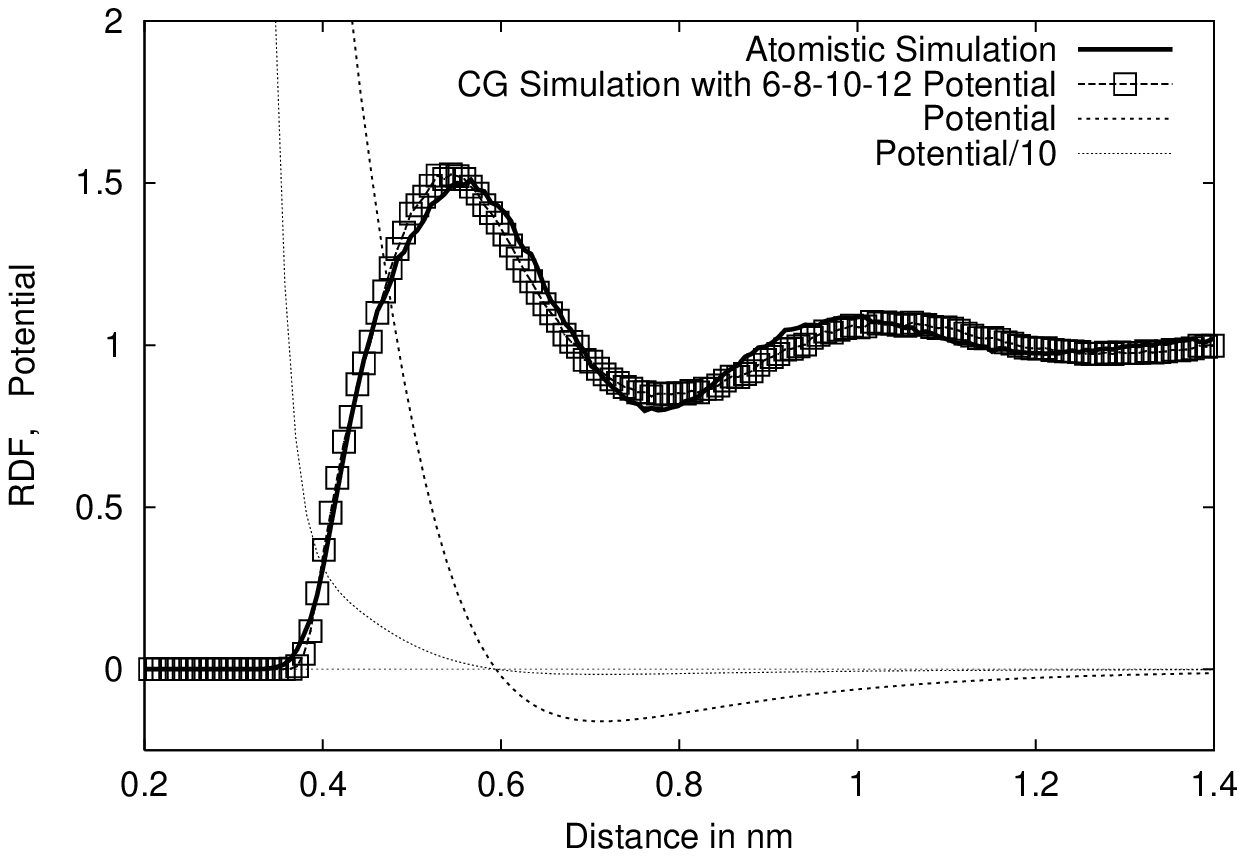}
    \caption[Modeling isoprene pseudo monomers with the 6-8-10-12
      potential: thick continuous line is the target center-of-mass RDF
      obtained from atomistic simulations. Squares represent the RDF
      obtained with the dotted potential. The thin dotted line is the
      potential divided by 10 to show the two repulsive regimes.]{}
    \label{f:rdfiso}
\end{figure}

As the first ``real'' system we discuss a liquid of isoprene pseudo
monomers.  The ultimate aim is to study trans-poly(isoprene) with one bead per
monomer.  In order to develop a coarse-grained potential for these
monomers, we performed atomistic simulations with 200 monomers at room
temperature ($T=300$ K, $p=1013$ hPa).  The force-field parameters of
these all-atom simulations were the same as derived for the simulation
of trans-poly(isoprene) in Ref.~\citen{faller:iso} where the internal
rotation parameters have been obtained by quantum chemistry
calculations.
  The pseudo monomers
are obtained by cutting the polymeric backbone at the double bonds
(dashed bond in Fig.\ \ref{f:mols}). The dangling bonds where the
polymeric backbone would go on are not considered in the simulation.  
The cut was placed at the double bond because the distribution
of bond lengths and bond angles of the resulting CG-chain are
sharper and exhibit less correlation than for a cut at the
CH$_2$--CH$_2$ bond. This is advantageous for the construction
of CG force fields of a polymer chain which is not further discussed
in this paper.
From the simulation of the atomistic monomer liquid we obtain a target
CM-RDF which is used in the following optimization (see thick continuous 
line in Fig.\ \ref{f:rdfiso}).
 It clearly shows a liquid structure with the first peak much
softer than that of a LJ liquid.

 The CG-simulations were performed with 400
spherical particles at constant volume using the same program as in
Sec.\ III.  The particle number density was fixed to match the density
of the atomistic simulations (817 kg/m$^3$).
The cutoff radius for the nonbonded
interactions was 1.26 nm.  In this case the 6-8-10-12 potential proved
to be successful (see Fig.~\ref{f:rdfiso}).  The final potential
exhibits two repulsive regimes: a very large slope below 0.35 nm marking
the hard core and a much smaller slope above this value which leads to
the broad first peak.
%
The successful potential was found with the parameterization
(\ref{e:potmm}) and has the parameters 
$\epsilon=9.3588$ kT, ${\mu_1}={\mu_2}=0.427$ nm, ${\mu_3}=0.738$
nm, and $\nu=0.5163$.  The simplex algorithm was restarted twice and
altogether about 200 simplex steps have been performed.
The pressure of the system with the best fitting RDF
is positive and about twice the pressure of the LJ6-12 reference system of
the preceding section.

This example demonstrates the technical capability of the algorithm to
generate a coarse-grained potential for a complex molecular fluid.
Physically, however, the obtained potential might be not very useful for
the simulation of polymers because of different densities of the polymer
melt and the pseudo-monomer melt due to overlap or end effects.  One
should also note that in the atomistic model the appearance of two
different repulsive regimes (which causes the broad first peak of the
RDF) is due to the oblate geometry of the molecule. It allows two
molecules to approach each other from different directions which have
different effective exclusion radii. This information is not contained
in the coarse-grained potential which is spherically symmetric. Hence,
even though the coarse-grained potential reproduces the RDF, it may be
inappropriate when the anisometry of the molecule is important. Similar
caveats hold for e.g.\ the model of THF discussed in the next paragraph.

\subsection{Tetrahydrofurane (THF)}
\begin{figure}
  \psfigure{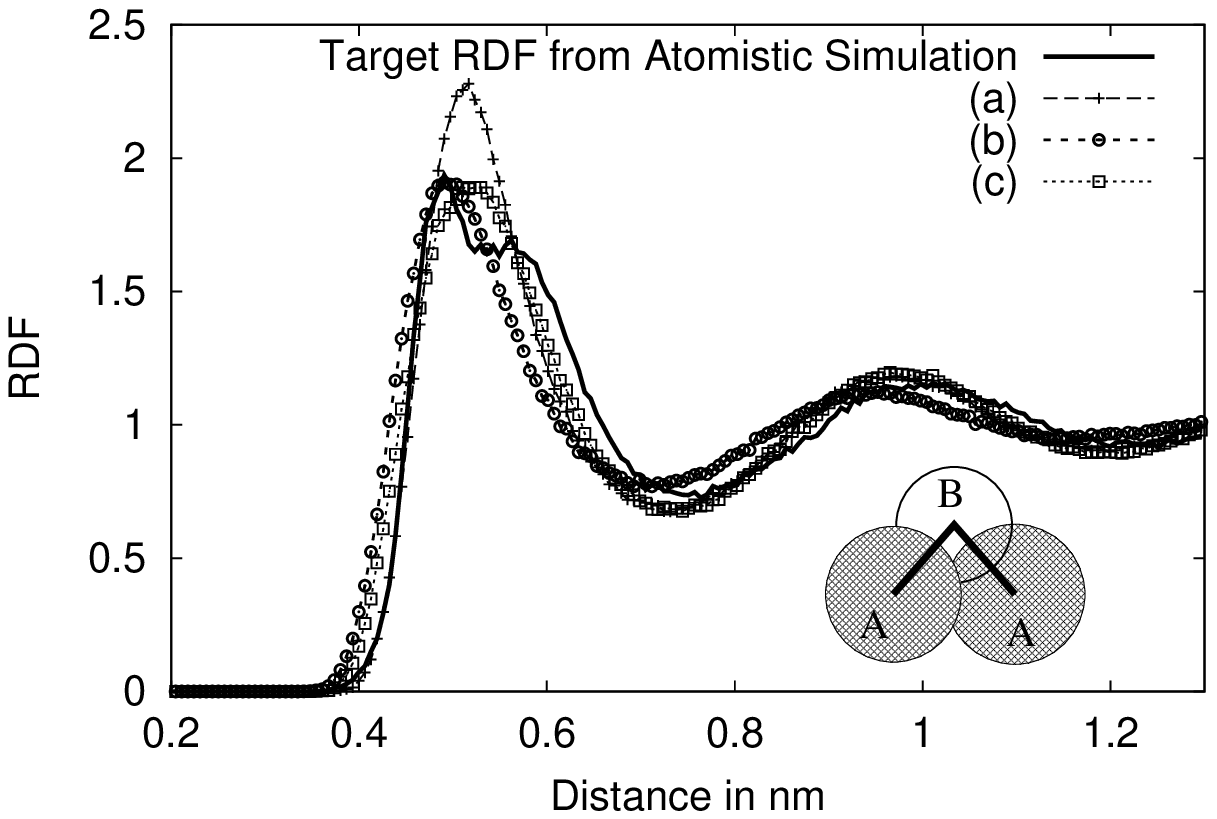}
  \caption[Modeling THF by three LJ6-12 spheres connected with bond
    length 0.3 nm and bond angle 120$^\circ$.
    Thick continuous line: target RDF from atomistic simulations.
    Examples of CG optimization steps:
 (a) $\epsilon_A=0.535$, $\sigma_A=0.425$ nm,
$\sigma_B=0.3625$ nm, and $\epsilon_B/\epsilon_A=1.525$;
 (b) $\epsilon_A=0.5344$, $\sigma_A=0.3469$ nm,
$\sigma_B=0.4234$ nm, and $\epsilon_B/\epsilon_A=1.584$;
 (c) $\epsilon_A=0.508$, $\sigma_A=0.3372$ nm,
$\sigma_B=0.496$ nm, and $\epsilon_B/\epsilon_A=1.307$.
   The features of the first peak are not reproduced.]{}
  \label{f:rdfthf2}
\end{figure}

The common solvent tetrahydrofurane (THF), 
for which an atomistic force field was developed in Ref.~\citen{faller99},
is proposed as a second example of a molecular liquid.  The target RDF
was obtained with the final force field from the simulation of that
reference (ambient conditions: $T=296$ K, $p=1013$ hPa; density
$\rho=886$ kg/m$^3$).
This CM-RDF has almost the form of a simple fluid's RDF except the small
tip on the first shell peak (see the thick line in
Fig.~\ref{f:rdfthf2}).  Careful examination of the atomistic data shows
that this tip can be attributed to some fraction of T-shaped
configurations of two neighboring THF molecules in contrast to a
parallel alignment. However, the center of mass distances for different
relative orientations are only slightly larger and do not result in a
detached second peak.  Our question was again if one can design a
coarse-grained model with simple spheres which reproduces this CM-RDF.
Obviously, a standard LJ6-12 potential will not work since there is no
structure in the potential minimum.  So we started with the 6-8-10-12
potential which can yield some structure on the first shell peak.
However, the produced peaks are too far away of each other to reproduce
the tiny feature at $r=0.5$ nm on the CM-RDF of THF (not shown).  After
lengthy trials, it turned out that potential (\ref{e:potmm}) is not
adapted to this problem.

We next tried to model the THF molecule by three LJ6-12 beads to account
for its slightly disk-like structure. The beads are connected
by harmonic bonds with a quite stiff spring constant, a harmonic angle
potential with the equilibrium angle 120$^\circ$, and their excluded volume is
strongly overlapping (inset in Fig.~\ref{f:rdfthf2}).
Two beads are supposed to be identical, the third one might have
different parameters ($\sigma$ and $\epsilon$). We chose the following
four parameters for the simplex optimization: $\epsilon_A$, $\sigma_A$,
$\sigma_B$, and $\epsilon_B/\epsilon_A$.
Fig.~\ref{f:rdfthf2} shows some typical RDFs obtained during
this CG-optimization which are not much better than the RDFs from
the 6-8-10-12-potential. They always reproduce  only some feature
of the target RDF, e.g.\ the starting slope of the RDF,
the position of the first sharp peak, the position of the
second half of the first peak, or an average of the first peak.
Often, the simplex optimization is stuck with a purely
repulsive potential.
Some sharper structure could be obtained on the first peak, but the
corresponding samples were frozen in an amorphous state with enormous
pressure (about 10 times the pressure of the LJ6-12 reference system).

\begin{figure}
  \psfigure{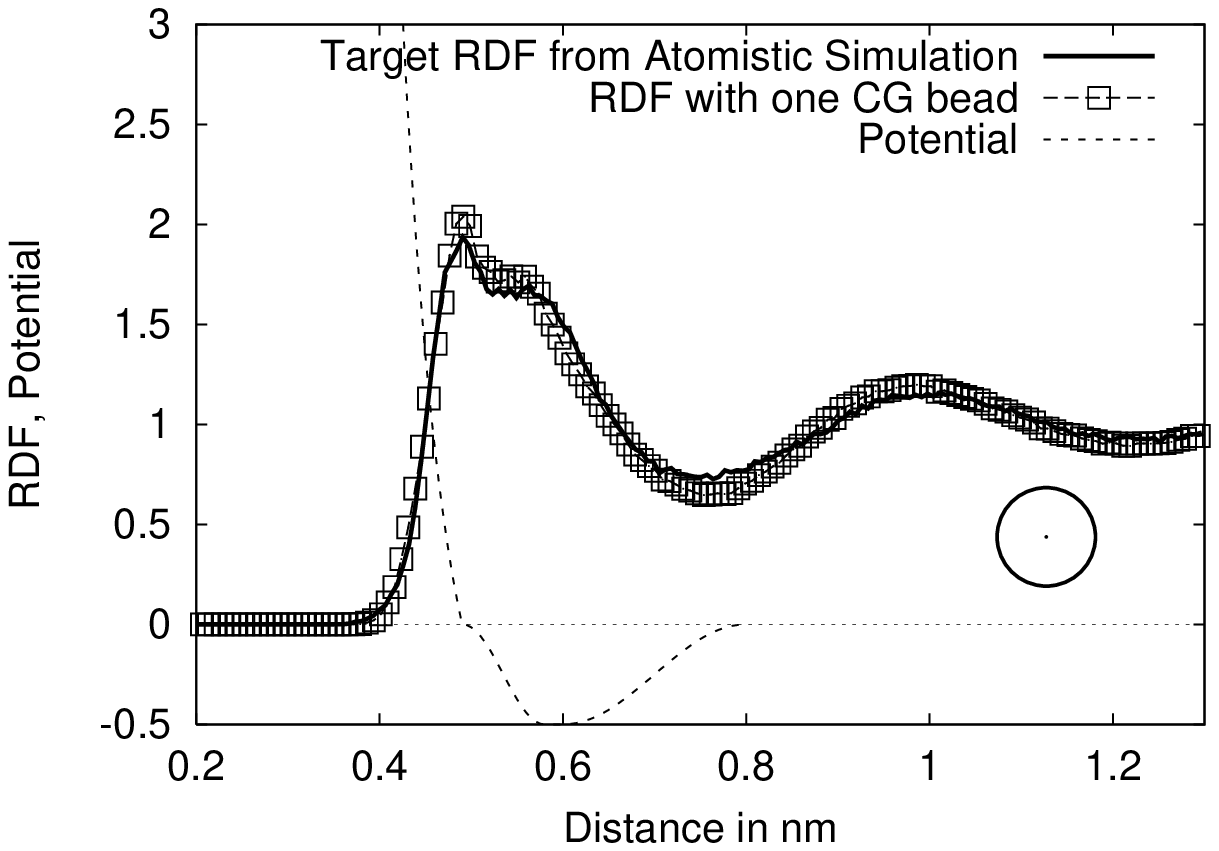}
  \caption[Modeling THF by one sphere with the piecewise defined
    potential Eq.~(\ref{e:potpiece}) and $\epsilon_1=1.7$ kT, $\sigma_1=\sigma_2=0.49$ nm, $\epsilon_2=0$,
$\epsilon_3=\epsilon_4=0.25$ kT, $\sigma_3 = 0.6$ nm, $\sigma_{\rm
  cut}=0.8$ nm.
    Thick continuous line: target RDF from atomistic simulations,
    broken line: potential, squares: RDF from CG simulation.]{}
  \label{f:rdfthf5}
\end{figure}

In principle, the 6-8-10-12 potential for one-bead models is able to
have a saddle point or a second minimum which should lead to at least a
shoulder in the RDF. However, this potential is still not of sufficient
generality to account for the location of the extrema and the potential
value as well as of the width of the extrema.  This is not only a
problem of our parameterization but of the potential form in general
where the first minimum is always much narrower than the second one due
to the $1/r^6$ factor  dominating at small distances.
This led us to developing a piecewise defined potential by
modeling the potential minimum by a sequence of different cosine
terms.
Eq.~(\ref{e:potpiece}) gives the form which was finally successful
with the following parameters for THF:
$\epsilon_1=1.7$ kT, $\sigma_1=\sigma_2=0.49$ nm, $\epsilon_2=0$,
$\epsilon_3=\epsilon_4=0.25$ kT, $\sigma_3 = 0.6$ nm, $\sigma_{\rm
  cut}=0.8$ nm.
The potential and the corresponding RDF are shown in
Fig.~\ref{f:rdfthf5}. With this potential, THF could be described by one
single bead. The pressure of this system is positive and only slightly larger
than in the LJ6-12 reference system.

\subsection{Diphenyl carbonate (DPC)}

Diphenyl carbonate (DPC) is a test case for a complex liquid of highly
anisotropic molecules.
Its CM-RDF is taken from atomistic simulations of Ref.~\citen{MeHaMP99}
at 393 K.
In all optimization steps, we fixed the particle number density
to the value of the atomistic simulations which is
3.04 molecules per nm$^3$.

\begin{figure}
    \psfigure{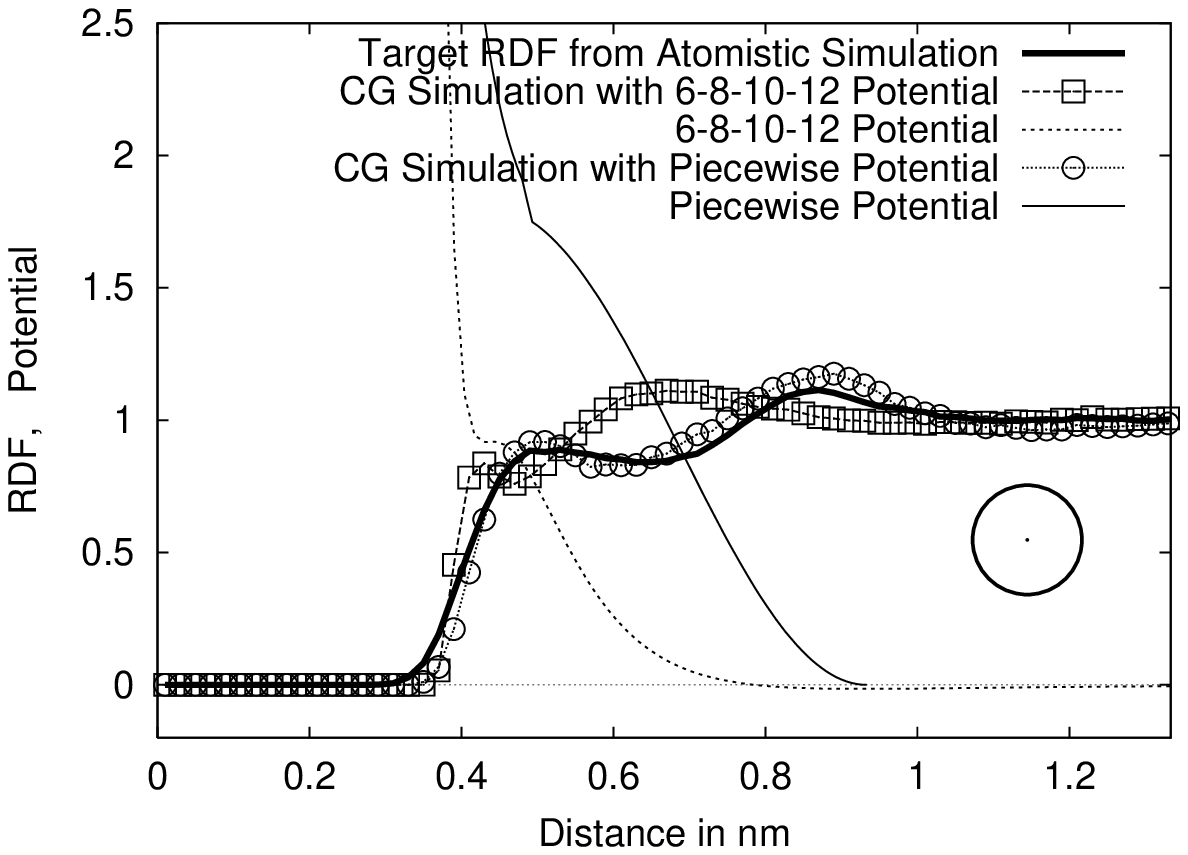}
    \caption[Modeling DPC by one spherical bead with a 6-8-10-12
      potential (dashed line) or a piecewise defined potential (thin
      continuous line).  The thick continuous
      line is the target RDF from atomistic all-atom simulations.
      Squares represent the RDF of spherical beads with the 6-8-10-12
      potential, circles the RDF corresponding to the piecewise potential.]{}
    \label{f:rdfdpc1}
\end{figure}

The simplest approximation is to model the DPC molecule by one spherical
bead. One might think that this cannot work at all, since the molecule
is really not spherical, but, using the 6-8-10-12 potential in the
parameterization (\ref{e:potmm}), one can at least reproduce the
qualitative features as shown in Fig.~\ref{f:rdfdpc1} (squares). The
corresponding potential is also shown as dotted line: it exhibits a
shoulder for the first small peak. The parameters of the shown potential
are $\epsilon=0.5883$ kT, ${\mu_1}=0.432$ nm, ${\mu_2}=0.4422$ nm,
${\mu_3}=0.927$ nm, and $\nu=0.5$. In this optimization run, we fixed
$\mu_1$ and $\nu$.
We performed about 20 restarts of the simplex algorithm with different
start values and different sets of parameters fixed but we could not get
a result with more than only qualitative coincidence of the RDFs.
This leads us to the conclusion that the parameterization
(\ref{e:potmm}) is not suited for producing a larger spacing between the
first two peaks.  Trying potentials with the second minimum at larger
distances, the particles stay where they are because they are pushed
into the repulsive regime due to the fixed volume.  This problem arises
because the 6-8-10-12 potential is not flexible enough.
As one can see already in Fig.~\ref{f:rdfdpc1}, a solution is found with
the piecewise potential already used in the preceding section. It allows
a much softer core. This makes it possible to reproduce the CM-RDF with
only one spherical bead much better. In this case, we defined a purely
repulsive potential with the parameters $\epsilon_1=0.38$ kT,
$\sigma_1=\sigma_2=0.5$ nm, $\epsilon_2=\epsilon_4=0$, $\epsilon_3=0.9$
kT, $\sigma_3 =\sigma_{\rm cut}=0.933$ nm.  The pressure in this system
is about 50\%\ larger than for the 6-8-10-12 potential plotted in the
same figure.

\begin{figure}
    \psfigure{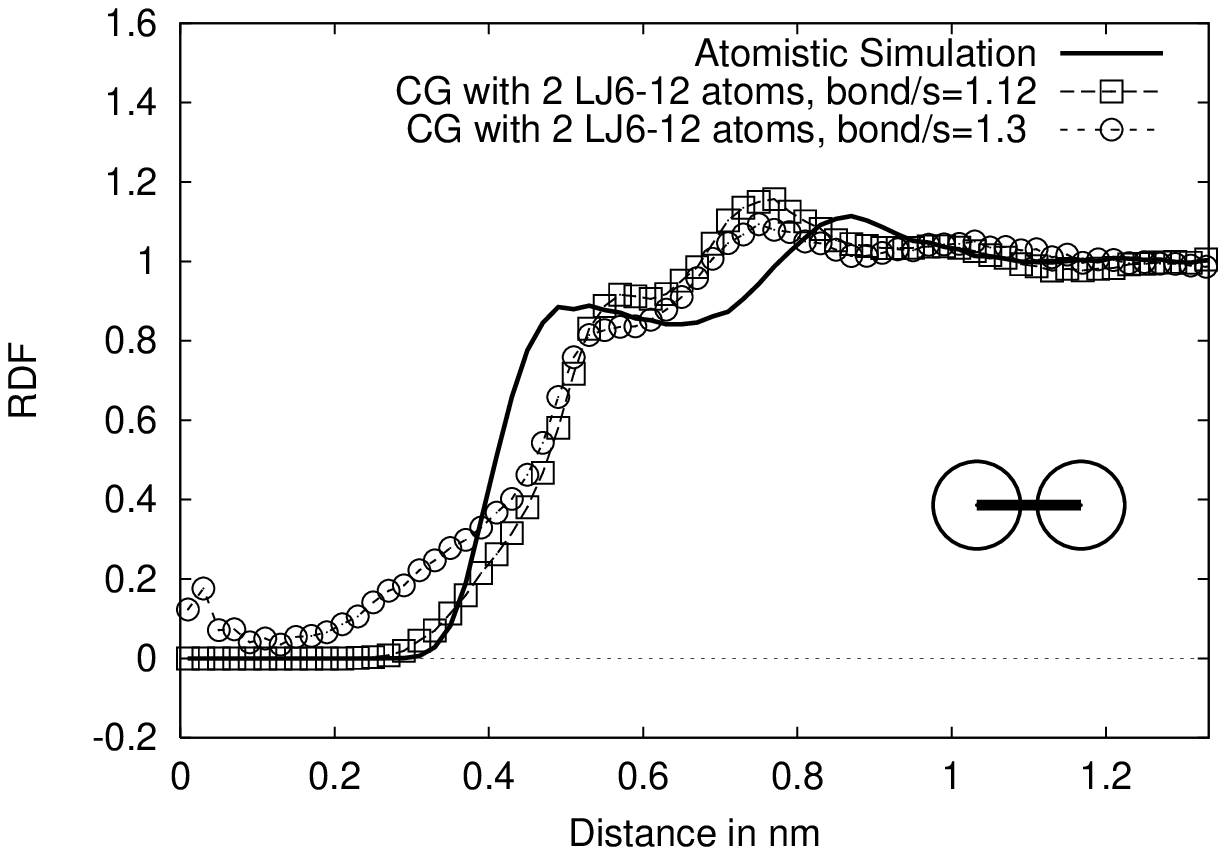}
    \caption[Modeling DPC by two LJ6-12 spheres connected
      with bonds of 0.6 nm (squares) and 0.667 nm (circles).
      The corresponding LJ-$\sigma$ of the CG beads are $0.533$ nm and
      0.517 which corresponds to a ratio bond length over LJ-$\sigma$
      of 1.12 and 1.3, respectively. In the second case, the spacing
      between the two beads forming a molecule is so large that other
      molecules can pass through.]{}
    \label{f:rdfdpc2}
\end{figure}

\begin{figure}
    \psfigure{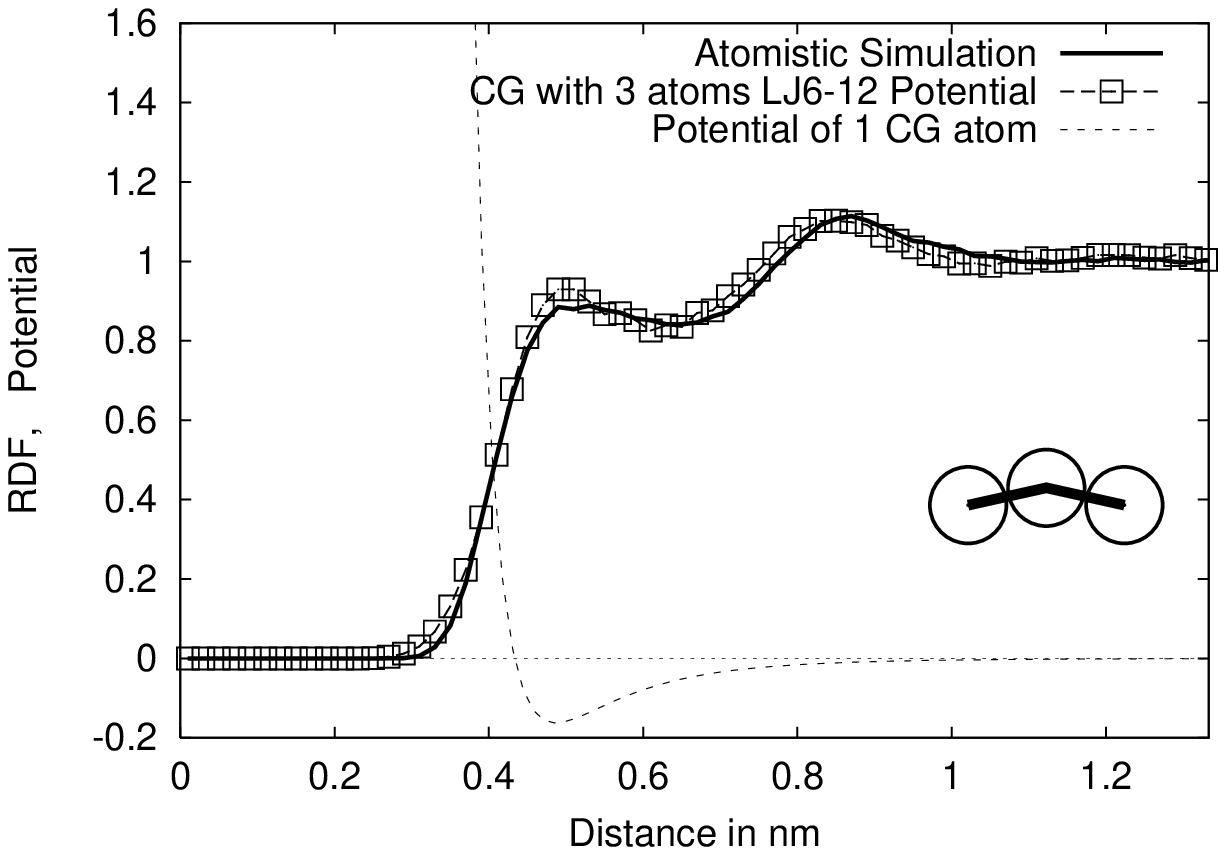}
    \caption[Modeling DPC by three LJ6-12 spheres which are connected
      with a bond length of 0.43 nm and a bond angle of 170$^\circ$. The
      optimized potential of a CG atom has LJ-$\sigma$=0.43 nm and
      LJ-$\epsilon$=0.16 kT for all three beads.]{}
    \label{f:rdfdpc3}
\end{figure}

To improve the model, we used a two-bead molecule which seems to be a
more natural attempt for modeling DPC. The two beads are connected by a
harmonic bond with a force constant allowing 10\%\ variation of bond length.
We performed optimization of the LJ-$\sigma$  with different bond
lengths.
 This yields the qualitative features of the pre-peak shoulder, but, again,
the distance of the two peaks is too small as one can see in
Fig.~\ref{f:rdfdpc2} (squares).
To get a larger spacing of the two peaks, we needed to increase the bond
length between the two CG beads. However, the bond length had to be so
long that molecules could pass through each other.  This yields an
unwanted and unphysical peak in the center of mass RDF at $R=0$ (circles
in Fig.~\ref{f:rdfdpc2}).
To avoid this, we inserted a third particle to enforce some excluded
volume in the center of the molecule.  During optimization, the
LJ-$\sigma$ of the inner bead was allowed to have another value than for
the two outer beads. For the bond between an outer and an inner bead
lengths from 0.33 to 0.43 nm were tried. To keep the three beads aligned
we added a stiffening potential with its equilibrium at 170$^\circ$ and
force constant $k_\theta=28$ kT/rad which corresponds approximately to
the angle distribution found in the atomistic simulations.  This,
eventually, yields a good correspondence of the RDFs as shown in
Fig.~\ref{f:rdfdpc3}. The pressure in the shown multi-bead systems
is about twice as high as in the one-bead models of DPC.
We started several optimization runs with different bond length.  The
free parameters during these optimization runs were the LJ-$\epsilon$
and the LJ-$\sigma$ for the inner and the outer beads.  The best fit was
found for bond length of 0.43 nm and almost the same LJ-sigma for all
three beads. Looking directly at the molecular dimensions, one gets some
different values: The distance from the phenyl ring centers to the
central carbon is 0.36 nm and the diameter of a phenyl ring is
approximately 0.6 nm. This difference is not surprising since the phenyl
rings are not spherical but more oblate objects.  The optimization shows
that this effect on the CM-RDF is best reproduced by smaller spheres at
a larger distance.
The perfect agreement of the RDFs in Fig.~\ref{f:rdfdpc3}
therefore shows that the peculiar features of the CM-RDF of DPC
simply follow from its shape.

\section{Discussion and Conclusion}

We reported in this work the construction of coarse-grained models of
molecules which are optimized with respect to their radial
distribution function. We implemented an automatic
optimization scheme using the simplex algorithm which was applied to
several complex liquids.
Several conclusions can be drawn.  The simplex algorithm is a useful tool
for finding and optimizing coarse-grained models against certain target
quantities. Not only single numbers but whole distribution functions can
be used as target quantity.  However, the automatic procedure
still needs a lot of supervision. 

 A good first guess of the adjustable
parameters is needed, otherwise the algorithm is trapped in local minima
which often are not of acceptable quality.  The method works best if one
has good start parameters.  If there is no reasonable guess from
the construction of the coarse-grained geometry,
one possibility to obtain such start
parameters is to fit the test potential to the potential of mean force
$-k_BT\ln g(r)$.  However, this works only for sufficiently simple
liquids. In fact, the potential of mean force is the result
of pair interactions as well as of many-particle interactions
of higher order. As the latter are not negligible,
the potential of mean force used as approximation of a pair potential
will not lead to satisfying results. This is especially the case
for  aspherical molecules as THF and DPC where the structure
of the CM-RDF is due to some directional interactions which
are intentionally omitted in the coarse-grained model.
Some interesting aspects to this context can also be found in Refs.\ 
\citen{Sop96cp} and \citen{Sop97jpcm} where Soper developed an empirical
potential MC method to derive the atomic positions from neutron
scattering data.  It is based on a self-consistent iteration starting
from the potential of mean force resulting in an empirical numerical
potential.  This approach might be also useful for deriving
coarse-grained simulation potentials.  In this work, we still stuck to
analytical potentials. However, with the piecewise potential of Eq.\ 
(\ref{e:potpiece}) we gave up any theoretical interpretation which could
not also be found in a purely numerical potential.

To avoid being trapped in local minima one could use global
optimizers like simulated annealing or genetic algorithms with tabulated 
potentials. This, however, is computationally much more demanding.
On the other hand, it is often possible  to come up with reasonable
guesses of the starting values based on knowledge of the system's
physics. Rewriting the potential expansion in terms of meaningful
parameters like in Eq.\ (\ref{e:potmm}) has already helped much in this
respect.
There are also more sophisticated forms or extensions of the simplex
optimization algorithm, e.g.\ parabolic extrapolation.  These
algorithms speed up the final convergence if one is already
sufficiently close to the optimum.  For our problem, they are not so
useful because, usually, the error bars on the RDF or other target
quantities are larger than the precision which could be gained by
further optimization steps where these algorithms become faster.

Concerning the physical problem addressed in this paper we conclude that
one can fit almost any molecular RDF with one particle and an appropriate
spherically symmetric potential.  Once one has abandoned conditions
about differentiability or certain ``nice'' functional forms, there are
no limits to the invention of potentials as Eq.~(\ref{e:potpiece}) to
reproduce funny RDFs.  However, restriction to spherically averaged
distributions might not be useful for the prediction of other
properties. Hence, for molecules with high asymmetry, the combination of
two or three spheres might be preferable to model structural properties.
The method could also be used to fit experimental RDFs similar to the
spirit of the `reverse MC' method.
The CG simulations of our examples where performed in the NVT ensemble.
The constant volume constraint imposes the correct density. 
For the transferability of the derived potentials, however, it will
often be useful to apply the pressure as an additional target
observable. This yields an additional term in the merit
function similar to  Ref.\ \citen{faller99}.

We have described a tool for optimizing nonbonded simulation
parameters, which, when used carefully, is very useful in force field
development. Several examples of complex liquids demonstrate how the
method works.  For the simulation of polymeric materials discussed in
the introduction, we are now working on optimizing polymer force
fields where nonbonded parameters are optimized by using oligomers
instead of single (pseudo)monomers.  For polymers, the connectivity is
another important aspect.  It is possible to adjust bonded
parameters as bond angle distributions of coarse-grained models by the
present approach. All this will be presented in a separate
publication.\cite{cgpaper2}

\section*{Acknowledgements}
We thank Markus Deserno, Kurt Kremer, and Heiko Schmitz
for many discussions and
Oliver Hahn for contributing an MD program.

\appendix
\section{Parameterizing the 6-8-10-12 potential}

Here we sketch a way of rewriting potentials such that
the parameters have an obvious interpretation.  The interesting points
of a potential are usually the locations of extrema or inflection points
and the function values at these points.  The locations are given by the
zeros of the derivative of the potential.  We show here the derivation
of the example of the 6-8-10-12 potential used in this paper. It has
nice symmetry since only even powers appear, but the idea is also
applicable to different expansions.

Rewrite the potential in the following way:
\begin{eqnarray} \label{e:pot1}
V(r) &=& {a_6 \over r^6} +{a_8 \over r^8}
 +{a_{10} \over r^{10}} +  {a_{12} \over r^{12}} \\
 &=& {\epsilon\over r^{12}} \left( r^6 + a r^{4} + b r^2 + c \right)
\end{eqnarray}
The derivative of $V(r)$ reads
\begin{equation}
V'(r) =  -{6\epsilon\over r^{13}} \left(r^6 + {8\over 6} a r^{4} +
  {10\over 6} b r^2 + {12\over 6} c \right)  \label{e:potderiv}
\end{equation}
We want to know the zeros of the polynomial in the parentheses.
Since it is complicated to solve a polynomial with degree
greater than two and in general impossible for degree greater than four,
we choose a constructive way: Be $\mu_i$ the zeros, then
$p(x) = \prod_i (x-\mu_i)$ is a polynomial having these zeros.
In our case, this gives
\begin{eqnarray}
p(x) &=& (r^2 - \mu_1^2)(r^2 - \mu_2^2)(r^2 - \mu_3^2) \\
     &=& r^6 - \left(\mu_{1}^{2}+\mu_{2}^{2}+\mu_{3}^{2}\right) r^{4}
         + \left(\mu_{1}^{2}\mu_{2}^{2}+\mu_{2}^{2}\mu_{3}^{2}
                 +\mu_{1}^{2}\mu_{3}^{2}\right) r^{2} 
         -  \mu_{1}^{2}\mu_{2}^{2}\mu_{3}^{2}   \label{e:polex}
\end{eqnarray}
Comparing the coefficients of the polynomial (\ref{e:polex}) with
eq. (\ref{e:potderiv}) one ends up with
\begin{eqnarray}
 a &=& - {6\over 8} \left(\mu_{1}^{2}+\mu_{2}^{2}+\mu_{3}^{2}\right) \\
 b &=&  {6\over 10}  \left(\mu_{1}^{2}\mu_{2}^{2}+\mu_{2}^{2}\mu_{3}^{2}
                 +\mu_{1}^{2}\mu_{3}^{2}\right) \\
 c &=& - {6\over 12}  \mu_{1}^{2}\mu_{2}^{2}\mu_{3}^{2}
\end{eqnarray}
which yields the potential form given in equation (\ref{e:potmm}).
$\pm\mu_{1}, \pm\mu_{2}, \pm\mu_{3}$ are the
locations of the extrema of $V(r)$, whereof only the positive locations
are of physical relevance.
In this parameterization, the relative height of the
extrema and the location of the minima are mutually dependent.
To have some more freedom, we chose the coefficient 0.5
of $c$ to be a supplementary variable $\nu$.


\clearpage
\listoftables

\begin{table}
\caption{Parameter sets of two optimization runs of the Buckingham exp-6
  potential (\protect\ref{e:exppara}) to fit the RDF of a LJ6--12 potential.}
\label{t:exp6}
\begin{tabular}{ldddddc}
$i$  &  $\epsilon$ & $\sigma$ & $b_1$   &  $b_2$  &   $f_i$  &    interval of fit \\
\hline
\multicolumn{6}{l}{(A) Start simplex} \\
\#1 &   668.0 & 0.4 & 13700.0 & 6.03 &  0.0082 &   0.8 - 1.5  \\
\#2 &   794.0 & 0.4 &  3404.0 & 5.45 &  0.0160 &   0.9 - 1.3  \\
\#3 &   840.0 & 0.4 &  1470.0 & 5.10 &  0.0401 &   1.0 - 1.2  \\
\#4 &   730.0 & 0.4 &  5635.0 & 5.60 &  0.1124 &   0.9 - 1.1  \\
\#5 &   624.0 & 0.4 & 17000.0 & 6.10 &  0.0065 &   0.8 - 1.2  \\
\multicolumn{6}{l}{ best value out of 100 steps:} \\
\#28 & 656.26 & 0.4 & 14214.5 & 6.026 & 0.0011 \\
\hline
\multicolumn{6}{l}{(B) Start simplex obtained by fit to 1/4 of the potential:} \\
\#1 &      163.0 & 0.4 & 14000.0 & 6.00 & 2.2401 &   0.8 - 1.5 \\
\#2 &      180.0 & 0.4 & 14000.0 & 6.04 & 2.3178 &   0.9 - 1.3 \\
\#3 &      182.0 & 0.4 & 15000.0 & 5.97 & 1.4783 &   1.0 - 1.2 \\
\#4 &      170.0 & 0.4 & 15500.0 & 6.10 & 2.6004 &   0.9 - 1.1 \\
\#5 &      200.0 & 0.4 & 16000.0 & 6.00 & 1.2725 &   0.8 - 1.2 \\
\multicolumn{6}{l}{best value out of 45 steps:} \\
\#35 &    238.73 & 0.4 & 15576.98 & 5.756 & 0.0359 \\
\end{tabular}
\end{table}
\begin{table}
\caption{Parameter sets of optimization runs of the LJ6--9 potential
  to fit the RDF of a LJ6--12 potential (in Eq.~\ref{e:potraw} 
  only parameters  $a_6$ and  $a_9$ are different from zero).
  The last column gives the ratio of the pressure with respect to
the pressure in the LJ6-12 reference system.}
\label{t:lj69}
\begin{tabular}{ldddd}
$i$  &  $a_6$ & $a_9$  &   $f_i$  & $p_i/p_{\rm LJ}$\\
\hline
\multicolumn{4}{l}{Start simplex (run1)} \\
\#1  &    -4.00 & 5.00 & 0.1371 & 2.4 \\ 
\#2  &    -4.60 & 4.00 & 1.2887 & 0.4 \\
\#3  &    -5.00 & 6.00 & 0.0792 & 2.2 \\
\multicolumn{4}{l}{best value of run1 (30 steps):} \\
\#12 &    -5.125 & 6.125 & 0.0775  & 2.3\\
\multicolumn{4}{l}{best value of run2 (11 steps):} \\
\#10 &    -5.137 & 6.080 & 0.0700  & 2.2\\
\multicolumn{4}{l}{best value of run3 (35 steps):} \\
\#16 &    -7.390 & 7.219 & 0.0315  & 0.68 \\
\multicolumn{4}{l}{best value of run4 (20 steps):} \\
\#17 &    -8.260 & 8.100 & 0.0283  & 0.55 \\
\end{tabular}
\end{table}

\begin{table}
\caption{Parameter sets of optimization runs of the 6-8-10-12
  potential (\protect\ref{e:potraw}) to fit the RDF of a LJ6--12 potential.}
\label{t:lj681012}
\begin{tabular}{ldddddc}
$i$  &  $a_6$ & $a_8$ & $a_{10}$   &  $a_{12}$  &   $f_i$ & \\
\hline
\multicolumn{5}{l}{(A) Start simplex} \\
\#1  &   $-$4.0 &  0.0 & 1.0 & 4.0 & 0.2384  \\
\#2  &    0.0 & $-$4.0 & 4.0 & 0.0 & 1.6349  \\
\#3  &   $-$0.5 &  0.0 & 0.0 & 0.1 & 11.8609  \\
\#4  &   $-$0.4 &  0.0 & 0.1 & 0.0 & 13.5487 \\
\#5  &   $-$0.4 &  0.0 & 0.1 & 0.1 & 7.8224 \\
\multicolumn{5}{l}{best points out of 110} \\
\#26 &   $-$3.064 & $-$0.119 & 0.859 & 3.002 & 0.0156 \\
\#51 &   $-$3.053 & 0.104 & 0.637 & 2.970 & 0.0123 \\
\#99 &   $-$3.044 & 0.046 & 0.646 & 2.971 & 0.0128  \\
\hline
\multicolumn{5}{l}{(B) Small variations of the LJ6-12 reference potential} \\
\#1   &    $-$4.0 &  0.0 & 0.10 & 4.0 & 0.0040 &  ** already perfect\\
\#2   &    $-$4.5 &  0.1 & 0.01 & 4.4 & 0.0227 \\
\#3   &    $-$3.5 & $-$0.2 & 0.00 & 3.9 & 0.0018 & ** already perfect \\
\#4   &    $-$4.1 &  0.1 & 0.20 & 4.1 & 0.0322 \\
\#5   &    $-$4.2 &  0.2 & 0.05 & 4.2 & 0.0320 \\
\end{tabular}
\end{table}

\clearpage
\listoffigures


\end{document}